\documentclass[12pt]{article}
\usepackage{amsmath}
\usepackage{graphicx,psfrag,epsf}
\usepackage{enumerate}
\usepackage{natbib}
\usepackage{textcomp}
\usepackage[hyphens]{url} 
\usepackage{hyperref}

\newcommand{\blind}{0}

\providecommand{\tightlist}{%
  \setlength{\itemsep}{0pt}\setlength{\parskip}{0pt}}

\usepackage{xcolor,soul}
\usepackage{xspace}

\begin{document}

\def\spacingset#1{\renewcommand{\baselinestretch}%
{#1}\small\normalsize} \spacingset{1}


\if0\blind
{
  \title{\bf Integrating data science ethics into an undergraduate
major: A case study}

  \author{
        Benjamin S. Baumer \thanks{Benjamin S. Baumer is Associate
Professor, Statistical \& Data Sciences, Smith College, Northampton, MA
01063 (e-mail:
\href{mailto:bbaumer@smith.edu}{\nolinkurl{bbaumer@smith.edu}}). This
work was not supported by any grant. The authors thank numerous
colleagues and students for their support.} \\
    Statistical \& Data Sciences, Smith College\\
     and \\     Randi L. Garcia \\
    Psychology and Statistical \& Data Sciences, Smith College\\
     and \\     Albert Y. Kim \\
    Statistical \& Data Sciences, Smith College\\
     and \\     Katherine M. Kinnaird \\
    Computer Science and Statistical \& Data Sciences, Smith College\\
     and \\     Miles Q. Ott \\
    Statistical \& Data Sciences, Smith College\\
      }
  \maketitle
} \fi

\if1\blind
{
  \bigskip
  \bigskip
  \bigskip
  \begin{center}
    {\LARGE\bf Integrating data science ethics into an undergraduate
major: A case study}
  \end{center}
  \medskip
} \fi

\bigskip
\begin{abstract}
We present a programmatic approach to incorporating ethics into an
undergraduate major in statistical and data sciences. We discuss
departmental-level initiatives designed to meet the National Academy of
Sciences recommendation for integrating ethics into the curriculum from
top-to-bottom as our majors progress from our introductory courses to
our senior capstone course, as well as from side-to-side through
co-curricular programming. We also provide six examples of data science
ethics modules used in five different courses at our liberal arts
college, each focusing on a different ethical consideration. The modules
are designed to be portable such that they can be flexibly incorporated
into existing courses at different levels of instruction with minimal
disruption to syllabi. We connect our efforts to a growing body of
literature on the teaching of data science ethics, present assessments
of our effectiveness, and conclude with next steps and final thoughts.
\end{abstract}

\noindent%
\vfill

\newpage
\spacingset{1.45} 

\newcommand{\R}{\textsf{R}\xspace}
\newcommand{\pkg}[1]{\texttt{#1}}
\newcommand{\miles}[1]{\sethlcolor{pink}\hl{[MO]: #1}}
\newcommand{\bert}[1]{\sethlcolor{blue}\hl{[AK]: #1}}
\newcommand{\katie}[1]{\sethlcolor{yellow}\hl{[KK]: #1}}
\newcommand{\ben}[1]{\sethlcolor{green}\hl{[BB]: #1}}
\newcommand{\randi}[1]{\sethlcolor{purple}\hl{[RG]: #1}}

\begin{quote}
``The potential consequences of the ethical implications of data science
cannot be overstated.''

---\citet{national2018data}
\end{quote}

\hypertarget{introduction}{%
\section{Introduction}\label{introduction}}

Data ethics is a rapidly-developing yet inchoate subfield of research
within the discipline of data science,\footnote{For example, the Data
  Science department at Stanford University lists ``Ethics and Data
  Science'' as one of its research areas:
  \url{https://datascience.stanford.edu/research/research-areas/ethics-and-data-science}.}
which is itself rapidly-developing \citep{nas2019roundtable}. Within the
past few years, awareness that ethical concerns are of paramount
importance has grown. In the public sphere, the Cambridge Analytica
episode \citep{rosenberg2018cambridge} revealed how the large scale
harvesting of Facebook user data without user consent was not only
possible, but permissable and weaponized for political advantage
\citep{davies2015cambridge}. Facebook CEO Mark Zuckerberg initially
characterized ``the idea that fake news on Facebook influenced the
{[}2016 United States Presidential{]} election in any way'' as ``pretty
crazy''---comments he later regretted \citep{levin2017zuckerberg}.
Nevertheless, the subsequent tongue-lashing and hand-wringing has led to
substantive changes in the policies of several large social media
platforms, including several prominent public figures being banned.
Popular books like \citet{o2017weapons}, \citet{eubanks2018automating},
\citet{noble2018oppression}, \citet{fry2018hello}, and \citet{d2020data}
have highlighted how
\href{https://en.wikipedia.org/wiki/Algorithmic_bias}{algorithmic bias}
(when automated systems systematically produce unfair outcomes) can
render even well-intentioned data science products profoundly
destructive.

These incidents have revived a sense among tech professionals and the
public at-large that ethical considerations are of vital importance. In
light of this, it seems clear that indifference to ethics in data
science is not an informed position.

As academics, it is our responsibility to educate our students about
ethical considerations in statistics and data science \emph{before} they
graduate \citep{utts2021enhancing}. To that end, recent work by
\citet{elliott2018teaching} addresses \emph{how} to teach data science
ethics. The machine learning community convenes the
\href{http://www.fatml.org/}{ACM Conference on Fairness, Accountability,
and Transparency} (which includes Twitter as a sponsor), which focuses
on ethical considerations in machine learning research and development.
The \href{https://ainowinstitute.org/}{AI Now Institute} at New York
University publishes research and policy resources surrounding the use
of artificial intelligence and algorithmic accountability. Some of the
first wave of data science textbooks include chapters on ethics
\citep{baumer2021mdsr}.

Most specifically, the National Academies of Sciences, Engineering, and
Medicine Roundtable on Data Science Postsecondary Education devoted one
of its twelve discussions to ``Integrating Ethics and Privacy Concerns
into Data Science Education'' \citep{nas2019roundtable}.
\citet{national2018data} includes the following recommendations for
undergraduate programs in data science:

\begin{quote}
Ethics is a topic that, given the nature of data science, students
should learn and practice throughout their education. Academic
institutions should ensure that ethics is woven into the data science
curriculum from the beginning and throughout.
\end{quote}

\begin{quote}
The data science community should adopt a code of ethics; such a code
should be affirmed by members of professional societies, included in
professional development programs and curricula, and conveyed through
educational programs. The code should be reevaluated often in light of
new developments.
\end{quote}

\noindent  In the major in statistical and data sciences at Smith
College, we have incorporated discussions of ethics (in one form or
another) into all of our classes, including the senior capstone, in
which about 25\% of the content concerns data science ethics. The
default position of indifference prevalent in the tech community is
exactly the problem we are trying to help our students recognize and
solve. In this sense, indifference to ethics in data science is counter
to the mission of our program, and in a larger sense to our profession.

Especially in light of concerns about academic freedom, we wish to
stress that this treatment is not about indoctrinating students about
\emph{what} to think, but rather to force students to grapple with the
often not-so-obvious ramifications of their data science work and to
develop their own compasses for navigating these waters
\citep{heggeseth-sdss}. It is not a political stance---it is an
educational imperative, as stressed by recommendations 2.4 and 2.5 in
\citet{national2018data}.

In this paper, we present a programmatic approach to incorporating
ethics into an undergraduate major in statistical and data sciences. In
Section \ref{review}, we review and delineate notions of ethics in data
science. We discuss departmental-level initiatives designed to meet the
NAS recommendation for integrating ethics into the curriculum from
top-to-bottom, and from side-to-side as well through co-curricular
programming in Section \ref{dept}. In Section \ref{patches} we provide
six different modules that focus on data science ethics that have been
incorporated into five different courses. The modules are designed for
portability and are publicly available at our website. \footnote{\url{https://bit.ly/2v2cf8n}}
We review evidence of our progress in Section \ref{assess}. Section
\ref{conclusion} concludes the paper with next steps and final thoughts.

\hypertarget{review}{%
\section{Ethical considerations in statistics and data
science}\label{review}}

Ethical considerations in statistics have been taught for decades, going
back to the classic treatment of misleading data visualization
techniques in \citet{huff1954lie}. However, there are additional nuances
to ethical concerns in data science. In defining ``data ethics,''
\citet{floridi2016what} propose that:

\begin{quote}
data ethics can be defined as the branch of ethics that studies and
evaluates moral problems related to data (including generation,
recording, curation, processing, dissemination, sharing and use),
algorithms (including artificial intelligence, artificial agents,
machine learning and robots) and corresponding practices (including
responsible innovation, programming, hacking and professional codes), in
order to formulate and support morally good solutions (e.g.~right
conducts or right values). This means that the ethical challenges posed
by data science can be mapped within the conceptual space delineated by
three axes of research: the ethics of data, the ethics of algorithms and
the ethics of practices.
\end{quote}

In this section, we review the literature on teaching data science
ethics under the three categories outlined by \citet{floridi2016what}.
While the distinctions made by \citet{floridi2016what} are helpful, we
use the term ``data science ethics'' to encompass the full suite of
subjects defined therein. Often, it is the interplay and dependence
between two or more of these categories that provide the richest ethical
dilemmas. Our discussions, especially in Section \ref{conclusion}, touch
upon possible inter-divisional synergies in the teaching of data science
ethics in a liberal arts context. That said, while a liberal arts
setting may provide some advantages, much of what we propose should be
portable to other types of institutions (see Section \ref{portable}).

\hypertarget{practices}{%
\subsection{Ethics of practices}\label{practices}}

From a legal perspective, the General Data Protection Regulation
\citep{gdpr}---which became enforceable in 2018---provides Europeans
with greater legal protection for personal data stored online than is
present in the United States. This discrepancy highlights the
distinction between ethical and legal considerations---the former should
be universal, but the latter are patently local. At some level, laws
reflect the ethical values of a country, but a profession cannot
abdicate its ethical responsibilities to lawmakers. As O'Neil notes:
``it is unreasonable to expect the legal system to keep pace with
advances in data science.'' \citep{nas2019roundtable} This is not to say
that government agencies are not involved. The United Kingdom now offers
guidance to practitioners via their
\href{https://www.gov.uk/government/publications/data-ethics-framework/data-ethics-framework-2020}{Data
Ethics Framework}. For oversight, Germany is considering recommendations
for a data science ethics review board \citep{tarran2019german}.

Major professional societies, including the American Statistical
Association (ASA) \citep{asa-ethics}, the Association for Computing
Machinery (ACM) \citep{acm2018ethics}, and the National Academy of
Sciences (NAS) \citep{national2009being}, publish guidelines for
conducting research. These documents focus on topics like
professionalism, proper treatment of data, negligence, and conflicts of
interest. Similarly \citet{tractenberg2019strengthening},
\citet{tractenbergteaching}, and \citet{gunaratnaethical} explore ethics
in statistical practice but don't mention newer concepts like
\emph{algorithmic bias}. \citet{eds2018mason} focuses on industry and
identifies five framing guidelines for building data products: consent,
clarity, consistency, control, and consequences. Their related blog post
\citep{loukides2018oaths} promotes the use of checklists over oaths, and
is the inspiration for the command line tool
\href{https://deon.drivendata.org/}{\texttt{deon}}.
\citet{canney2015framework} present a framework for evaluating ethical
development in engineers. \citet{washington2020whose} examine how these
ethical codes often protect the interests of corporations and
professional associations at the expense of vulnerable populations.

A broader discussion of professional ethics in statistics and data
science would include issues surrounding reproducibility and
replicability, which would in turn include concepts like transparency,
version control, and p-hacking
\citep[\citet{wasserstein2019moving}]{wasserstein2016asa}. What is more,
inappropriate statistical analysis remains a problem in many fields,
including biostatistics \citep{wang2018researcher}.

\hypertarget{ethics-of-data}{%
\subsection{Ethics of data}\label{ethics-of-data}}

Within statistics, a major ethical focus has been on human subjects
research. \emph{The Belmont Report} is still required reading by
institutional review boards (IRBs) \citep{department2014belmont}. It
posits three major ethical principles (respect for persons, beneficence,
and justice) and outlines three major applications (informed consent,
assessment of risks and benefits, and selection of subject). Yet just as
we reject the argument that all legal data science projects are ethical,
we question the supposition that all IRB-approved data science projects
are ethical. Many IRBs have not been able to keep pace with the rapid
development of data science research, and have little authority over
research fueled by data collected by corporations.

For example, Facebook data scientists manipulated the news feeds of
689,003 users in order to study their ``emotional contagion''
\citep{kramer2015pnas}. While Facebook did not break the law because
users relinquished the use of their data for ``data analysis, testing,
{[}and{]} research'' when they agreed to the terms of service, many
ethical questions were subsequently raised, notably whether informed
consent was legitimately obtained. Moreover, Cornell University IRB
approval was obtained only \emph{after} the data had been collected,
meaning that the approval covered the analysis of the data, not the
collection or the design of the experiment. This example illustrates how
many university IRBs are ill-equipped to regulate ``big data'' studies
\citep{robinson2014theatlantic}.

More modern manifestations of data ethics are brought on by ``big
data.'' These include ethical concerns when scraping data from the web,
storing your personal data online, de-identifying and re-identifying
personal data, and large-scale experimentation by internet companies in
what \citet{zuboff2018surveillance} terms ``surveillance capitalism.''

\hypertarget{ethics-of-algorithms}{%
\subsection{Ethics of algorithms}\label{ethics-of-algorithms}}

While the ethics of practices and of data remain crucially
important---and continue to play a role in our curriculum---much of our
focus is on examining the impact of deployed data science products. Most
notably, we center questions of algorithmic bias (which are not simply
reducible to the use of biased data). The machine learning community is
having intense debates about the extent to which data or algorithms are
ultimately most responsible for bias in facial recognition and other
AI-driven products \citep{synced2020ai}. The impact of data science
products upon people having marginalized identities
\citep{vakil2018ethics}, particularly with respect to race and gender
\citep{gebru2020race}, is a growing focus of inquiry. In addition to
bias, \citet{bender2021dangers} also raise questions about the
environmental impact of Google's large language models.

\hypertarget{data-science-ethics-broadly-construed}{%
\subsection{Data science ethics, broadly
construed}\label{data-science-ethics-broadly-construed}}

These ethical areas are obviously informed by longstanding ethical
principles, but are distinct in the way that computers, the Internet,
databases, and tech companies have transformed the way we live
\citep{hand2018aspects}.

Our focus areas mostly intersect with those identified by
\citet{national2018data} as needed by data scientists:

\begin{itemize}
\tightlist
\item
  Ethical precepts for data science and codes of conduct,
\item
  Privacy and confidentiality,
\item
  Responsible conduct of research,
\item
  Ability to identify ``junk'' science, and
\item
  Ability to detect algorithmic bias.
\end{itemize}

This paper offers examples for implementing these focus areas. For
example, Section \ref{capstone} contains a module that has students
apply ethical codes in context (ethics of practices). The modules in
Sections \ref{okcupid} and \ref{greys} explore notions of privacy and
confidentiality (ethics of data). Sections \ref{race} and \ref{greys}
provide modules that illuminate notions of responsibility when
conducting research (ethics of practices). Sections \ref{stitchfix} and
\ref{capstone} present modules that encourage students to detect
algorithmic bias in action (ethics of algorithms).

Yet we also go beyond these key areas. The module in Section \ref{music}
explores boundaries between legal and ethical considerations (ethics of
data). In other activities not presented here, we engage students in our
senior capstone and machine learning courses with deep questions about
the impact that actions by large-scale internet companies have on our
lives (ethics of data \emph{and} algorithms). Our program's conception
of data science ethics is broad and inclusive. Thus, while published
ethical guidelines can be used to inform ethical judgments, we encourage
our students to consider these guidelines as non-exhaustive, especially
given their nascent and rapidly evolving nature. One exercise in the
capstone course asks students to consider what ethical precepts are
missing from various ethical guidelines.

Table \ref{tab:modules} summarizes these modules and links them to the
relevant categories as defined by \citet{floridi2016what} and the
taxonomy of learning defined by \citet{bloom1956taxonomy}.

\begin{table}

\caption{\label{tab:modules}Summary of ethical modules described. Categories correspond to those identified by Floridi and Taddeo. `Bloom' refers to Bloom's taxonomy.}
\centering
\begin{tabular}[t]{c|l|l|l}
\hline
Section & Topic & Categories & Bloom\\
\hline
\ref{okcupid} & OkCupid & data & Application\\
\hline
\ref{stitchfix} & StitchFix & algorithms & Application\\
\hline
\ref{greys} & Grey's Anatomy & practices, data & Application\\
\hline
\ref{music} & Copywriting music & practices & Evaluation\\
\hline
\ref{race} & Coding race & practices, data & Synthesis\\
\hline
\ref{capstone} & Weapons of Math Destruction & practices, algorithms & Evaluation\\
\hline
\end{tabular}
\end{table}

\hypertarget{approaches}{%
\subsection{Approaches to teaching data science
ethics}\label{approaches}}

While discussion about data science ethics abounds, there are few
successful models for how statisticians and data scientists can teach it
\citep{schlenker2019ethics}. Indeed, relevant work on teaching data
science by \citet{donoho201750}, \citet{hicks2018guide},
\citet{baumer2015ads}, \citet{hardin2015ds}, and
\citet{kaplan2018teaching} barely mentions ethics if at all. The first
reading of criteria for accreditation in data science published by ABET
did not include ethics (although the second reading does)
\citep{blair2021establishing}. Despite recommending the inclusion of
ethics into data science curricula, even the \citet{national2018data}
report does not include explicit recommendations for \emph{how} to do
so. One of the primary challenges is that while educators are typically
well-trained in the ethics of human subjects research, few have specific
training in, say, algorithmic bias, or even general ethical philosophy.
But why should a lack of training prevent us from teaching our students?
As \citet{bruce2018five} points out, ethical issues are not really a
technical problem, but rather ``a general issue with the impact of
technology on society,'' to which we all belong. We might make up for
our lack of training by partnering with philosophers and ethicists to
develop a robust ethical curriculum \citep{bruce2018five}.

Echoing \citet{bruce2018five} that ``there is a long history of scholars
and practitioners becoming interested in ethics when faced with new
technologies,'' \citet{gotterbarn2018thinking} argue forcefully that the
recent uptick in interest in ``computing ethics'' is merely the most
recent star turn for a longstanding and valued component of the computer
science curriculum. While this is surely true at some level and
important to keep in mind, it hardly seems like the renewed attention on
ethics is unwarranted. Moreover, \citet{gotterbarn2018thinking}'s focus
is on artificial intelligence driven systems like self-driving cars,
whereas our focus is on ethical questions concerning data collected
about people.

Several examples of how to teach ethics in statistics, data science, and
(mostly) computer science exist. \citet{hoffmann2021teaching} summarize
efforts in computer science and engineering education, including
discussion of specific teaching strategies. \citet{neff2017critique}
takes a broad view of data science ethics, bringing tools from critical
data studies to bear on the practice of actually doing data science.
\citet{burton2018teach} outlines a strategy for teaching computer
science ethics through the use of science fiction literature.
\citet{elliott2018teaching} provides a framework for reasoning about
ethical questions through the dual prisms of Eastern (mainly
Confucianism) and Western ethical philosophies. We found this inclusive
approach to be particularly valuable given the large presence of
international (particularly Chinese) students in our classes. Perhaps
presaging many recent scandals, \citet{zimmer2010but} analyzes a
Facebook data release through an ethical lens.
\citet{chivukula2021surveying} and \citet{shapiro2020re} discuss
approaches to teaching data ethics through human-computer interaction.
Fiesler analyzes ethical topics in a variety of computer science courses
\citep{saltz2019integrating, fiesler2020we, skirpan2018ethics}.
\citet{grosz2019} describes how ethics education is integrated into the
computer science curriculum at Harvard. Barocas teaches an undergraduate
elective course on data science ethics at Cornell
\citep{nas2019roundtable}. The University of Michigan now offers a
``Data Science Ethics'' course through both
\href{https://www.coursera.org/learn/data-science-ethics}{Coursera} and
\href{https://www.edx.org/course/data-science-ethics}{edX}.
\citet{carter2019ethics} provide a set of ethics ``labs'' for computer
science that might complement the ones we present here. Through 2021,
the
\href{https://foundation.mozilla.org/en/what-we-fund/awards/responsible-computer-science-challenge/}{Mozilla
Responsible Computer Science Challenge} has awarded \$3.5 million for
the development of ``curricula that integrate ethics with undergraduate
computer science training.''

These articles illustrate the need to further advance the teaching of
data science ethics in different institutional contexts. In this paper,
we present six additional concrete modules for teaching data science
ethics. We also outline departmental initiatives for fully integrating
ethics into an undergraduate data science curriculum and culture.

\hypertarget{dept}{%
\section{Department-level initiatives}\label{dept}}

At Smith, every department periodically reviews and updates a list of
\href{(https://www.smith.edu/about-smith/institutional-research/learning-goals\#offices-institutional-research-statistical-data-sciences)}{learning
goals} for their major. The major in statistical and data sciences (SDS)
is designed to cover a broad range of topics to produce versatile future
statisticians and data scientists. Our learning goals include skills
like: fitting and interpreting statistical models, programming in a
high-level language, working with a wide variety of data types,
understanding the role of uncertainty in inference, and communicating
quantitative information in written, oral, and graphical forms. Most
recently, we added the following learning goal:

\begin{quote}
Assess the ethical implications to society of data-based research,
analyses, and technology in an informed manner. Use resources, such as
professional guidelines, institutional review boards, and published
research, to inform ethical responsibilities.
\end{quote}

In support of this learning goal, we have taken measures to:

\begin{itemize}
\tightlist
\item
  incorporate ethics into all of our classes, culminating in a thorough
  treatment in the senior capstone course;
\item
  support student engagement in extra-curricular and co-curricular
  events that touch on data science ethics;
\item
  bring a diverse group of speakers to campus to give public lectures
  that often focus on ethical questions;
\item
  include a candidate's ability to engage with data science ethics as a
  criterion in hiring;
\item
  increase inclusion at every level of our program.
\end{itemize}

\noindent We discuss six specific modules for courses in Section
\ref{patches}. In this section we discuss approaches for the other
measures. We recognize that not every institution has the curricular
flexibility and resources that we have at Smith, nor is our student body
representative of those at different types of institutions (e.g., R1's
or two-year colleges). We discuss some specific considerations related
to teaching data science ethics in a liberal arts context in Section
\ref{sec:next}. Nevertheless, most of the modules we present can fit
into a single class period, which should provide instructors at any
institution with a reasonable opportunity to incorporate some of this
material.

\hypertarget{student-engagement-in-ethics}{%
\subsection{Student engagement in
ethics}\label{student-engagement-in-ethics}}

Our students are very interested in ethical questions in data science
(see Section \ref{survey}). As digital natives, they bring an
importantly different perspective to questions about, for example,
sharing one's personal data online. Many of them have never seriously
considered the ramifications of this. The notion that ``if you're not
paying for the product, then you are the product'' is new, scary,
challenging, relevant, personal, and engaging to them in a way that
helps them see data science as more than just a battery of technical
skills \citep{fitzpatrick2010lifehacker}. Thus, teaching ethics in data
science is another way to foster student interest in the
\emph{discipline}. Framing ethical questions in data science as unsolved
problems helps students imagine themselves making meaningful
contributions to the field in a way that may seem too remote of a
possibility in, say, estimation theory.

In particular, algorithmic bias intersects with questions about
inclusion and diversity with which students are already grappling on a
daily basis. During the past few years, we have applied for (and
received) funds from the community engagement center and the Provost's
office to support student engagement with the \emph{Data for Black
Lives} conference \citep{d4bl}. In 2018, the first year of the Data for
Black Lives conference, we hosted a remote viewing party on campus. In
2019, one of us attended the conference with five students. This
experience led to \emph{a student} inviting Data for Black Lives founder
\href{https://smithcollege-sds.github.io/sds-www/speaker.html}{Yeshimabeit
Milner to campus for a public lecture entitled ``Abolish Big Data.''}
{[}Milner is not against the use of data per se, but identifies
unethical data science products powered by the large-scale collection of
user data as newfangled examples of longstanding systemic racism.{]}
These experiences help students connect what they are learning in the
classroom to larger movements in the real world, and give them the sense
that their skills might be used to affect positive change in the
world---a powerful motivator.

We are fortunate that our institution provides generous funding for
bringing outside speakers to campus, and we have taken full advantage of
their largesse over the past two years. We welcomed BlackRock data
scientist Dr.~Rachel Schutt to give a talk titled ``A Humanist Approach
to Data Science,'' in which she underscored the importance of
recognizing the people behind the numbers, and highlighted examples of
recently published research that raised profound ethical dilemmas.
Dr.~Terry-Ann Craigie of Connecticut College\footnote{Dr.~Craigie is now
  faculty at Smith College.} came to talk about the intersections of
race, data science, and public policy. Dr.~Emma Benn of Mount Sinai
discussed how her intersectional social identity has informed her work
as a biostatistician. Alumna Gina DelCorazon spoke about her experiences
as Director of Data \& Analytics at the National Math and Science
Initiative in her talk ``From Interesting to Actionable: Why good
context matters as much as good code.'' At the invitation of a student
group, Dr.~Alisa Ainbinder, an alumna working locally in data science,
discussed ethical considerations in her work in non-profit accelerator
programs. Hearing from professionals about the ethical considerations in
their work helps reinforce the messaging we give them in class.

\hypertarget{program}{%
\subsection{Programmatic efforts in ethics}\label{program}}

The SDS major at Smith includes an ``application domain'' requirement.
One of the purposes of this requirement is to ensure that students
understand that all data and analyses have a context. Conducting ethical
data analysis requires knowledge of the context in which the data are
being used. For example, only through having some understanding of the
history of racial/ethnic groups in the United States can data scientists
hope to code and use race appropriately in their analyses (see Section
\ref{race}).

The SDS major at Smith requires every student to take one course that
focuses explicitly on communication. Another simple initiative was to
allow students to fulfill this requirement by taking the ``Statistical
Ethics and Institutions'' course taught at nearby Amherst College by
Andreas V. Georgiou, the former President of the Hellenic Statistical
Authority \citep{langkjaer2017trials}.
\href{https://en.wikipedia.org/wiki/Andreas_Georgiou}{Georgiou} has
faced criminal charges in Greece for his role in reporting economic
statistics around the time of the Greek debt crisis. The American
Statistical Association---among other organizations---has defended
Georgiou's actions \citep{pearson2021georgiou}. Although the course did
not explicitly focus on communication, we made an exception to our
policy to allow students to have this unique opportunity to learn about
statistical ethics from the person at the center of a world-famous
episode. Moreover, ethics and communication are intertwined, in that
conveying ethical subtleties requires a different skill set than say,
explaining a statistical model.

Finally, we take small steps to ensure that incoming faculty are capable
of supporting our program in meeting this newest learning goal. They
cannot be dismissive of ethical concerns in data science. In the same
way that a candidate who didn't understand correlation would not be
hireable, we consider whether a candidate who seemed ignorant of data
science ethics would be hireable. To assess this, we might ask a
question about data science ethics during a first round or on-campus
interview. We might ask candidates to submit a separate statement on
data science ethics as part of their application, or to discuss ethical
considerations in their teaching and/or research statement. To be clear,
we cannot and do not infringe upon the candidate's academic freedom by
assessing \emph{what} they think about data science ethics. Rather, we
are merely trying to assess \emph{how deeply} they have thought about
data science ethics and thus whether they are sufficiently prepared to
help the program meet our learning goals.

\hypertarget{patches}{%
\section{Modules for teaching data science ethics}\label{patches}}

In this section we present six modules for teaching ethics in data
science that are used in a variety of courses. Here, we give a brief
description of each module, its learning goals, and the context of the
course in which it is delivered. In our supplementary materials, we
provide more complete teaching materials.

\hypertarget{okcupid}{%
\subsection{Three uses of OkCupid data}\label{okcupid}}

\href{http://www.okcupid.com}{OkCupid} is a free online dating service
whose data has been scraped on at least three known occasions.
\citet{kim2015okcupid} presented scraped data on nearly 60,000 OkCupid
users in the early 2010's for use in the classroom. Around that same
time, Chris McKinlay created 12 fake OkCupid accounts and wrote a Python
script that harvested data from around 20,000 women from all over the
country \citep{poulsen2014wired}. In 2016, \citet{kirkegaard2016okcupid}
published a paper in an open-access psychology journal investigating a
variety of hypotheses about OkCupid users---along with the corresponding
data from 70,000 users. From the same underlying data source, these
three incidents provide fertile ground for substantive discussions about
the corresponding ethical considerations.

Some further detail reveals fascinating disparities:

\begin{itemize}
\item
  \citet{kim2015okcupid} obtained explicit permission from OkCupid CEO
  Christian Rudder before publishing the data in a statistics education
  journal. Their goal was to illuminate statistical phenomenon using
  data that was relevant to undergraduate students. In addition, the
  authors removed usernames from the data as a modest attempt at
  de-identifying the users. Only later were the authors
  \href{https://github.com/rudeboybert/okcupiddata/issues/1}{alerted to
  the fact that} even though usernames had been stripped, the full-text
  of the \texttt{essay} field often contains personally-identifying
  information like Facebook and Instagram handles. The original
  publication was subsequently corrected in May 2021 following the
  suggestions of \citet{xiao2021letter} to the editor of the
  \emph{Journal of Statistics and Data Science Education}
  \citep{kim2015okcupidcorrection}.
\item
  McKinlay did not publish the data he collected---his goal was
  personal. Essentially, he trained his own models on the data he
  collected to find his own match. It worked---he is now engaged to the
  woman he met. Only after his story was published were
  \href{https://pando.com/2014/01/22/did-the-mathematician-who-hacked-okcupid-violate-federal-computer-laws/}{questions
  raised} about whether he had violated the
  \href{https://en.wikipedia.org/wiki/Computer_Fraud_and_Abuse_Act}{Computer
  Fraud and Abuse Act}.
\item
  \citet{kirkegaard2016okcupid} included username, age, gender, and
  sexual orientation in the data set. This meant that users were easily
  identifiable and particularly vulnerable. While the blowback in this
  case was immediate,
  \href{http://fortune.com/2016/05/18/okcupid-data-research/}{Kirkegaard
  insisted that the data were already public and his actions were
  legal}. (\citet{zimmer2010but} sheds light on a similar episode
  involving Facebook.)
\end{itemize}

Collectively, these episodes raise issues about informed consent, data
privacy, terms of use, and the distinction between laws and ethics. One
could use these incidents to motivate coverage of technical concepts
such as
\href{https://en.wikipedia.org/wiki/K-anonymity}{\(k\)-anonymity}
\citep{sweeney2002k} and
\href{https://en.wikipedia.org/wiki/Differential_privacy}{differential
privacy} \citep{dwork2006calibrating}. In our senior capstone course
(see Section \ref{capstone}), we ask students to break into three groups
and discuss the relevant ethical issues involved in each case. Then, we
bring students together to write a coherent response. Some students
elect to use these incidents as the subject of a longer essay, as
described in Section \ref{capstone}.

\hypertarget{stitchfix}{%
\subsection{Algorithmic bias in machine learning}\label{stitchfix}}

Discussions on the perniciousness of ``algorithmic bias'' in machine
learning and artificial intelligence have become more prevalent of late,
both in the news media as well as in academic circles
\citep{noble2018oppression, eubanks2018automating, o2017weapons}.
However, few of these ideas have been incorporated into the classroom.
For example, in \citet{james2013islr}---a popular introductory textbook
on machine learning---the \texttt{Credit} dataset is often used as an
example (it is available in the companion \texttt{ISLR} R package
\citep{R-ISLR}). Readers are encouraged to apply various predictive
algorithms to predict the credit card debt of 400 individuals using
demographic predictors like \texttt{Age}, \texttt{Gender} (encoded as
binary), and \texttt{Ethnicity} with levels \texttt{African\ American},
\texttt{Asian}, and \texttt{Caucasian}. While the data are simulated,
one must still wonder what kind of thinking in students are we tacitly
encouraging by using ethnicity to predict debt and thus perhaps credit
score. This is especially fraught in light of existing inequalities to
access to credit that fall on demographic lines.\footnote{The subsequent
  \texttt{ISLR2} \citep{R-ISLR2} update to the \texttt{ISLR} package
  removes the \texttt{Gender} and \texttt{Ethnicity} variables from the
  \texttt{Credit} dataset.} In other words, to quote \citet{d4bl},
``What are we optimizing?''

In this module, we propose a hands-on in-class activity to help students
question the supposed objectivity of machine learning algorithms and
serve as a gateway to discussions on algorithmic bias. The activity
centers around \href{https://www.stitchfix.com/}{StitchFix}, an online
clothing subscription service that uses machine learning to predict
which clothes consumers will purchase. New users are asked to complete
either a men's or women's ``Style Profile'' quiz, whose responses are
then used as predictor information for the company's predictive
algorithms. However, both quizzes differ significantly in the types of
questions asked, how the questions are asked, in which order they are
asked, and what information and visual cues are provided.

\begin{figure}

{\centering \includegraphics[width=0.49\linewidth]{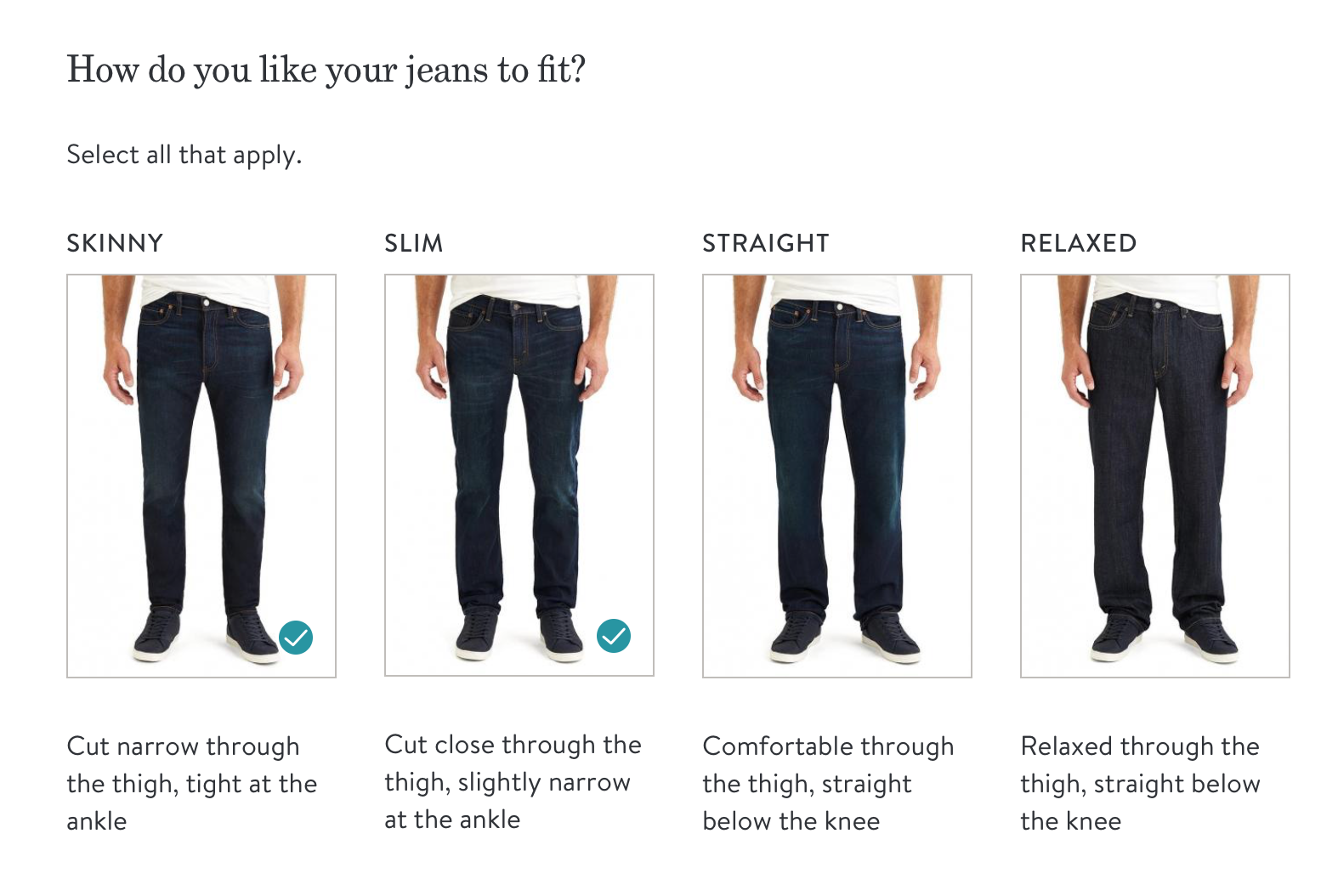} \includegraphics[width=0.49\linewidth]{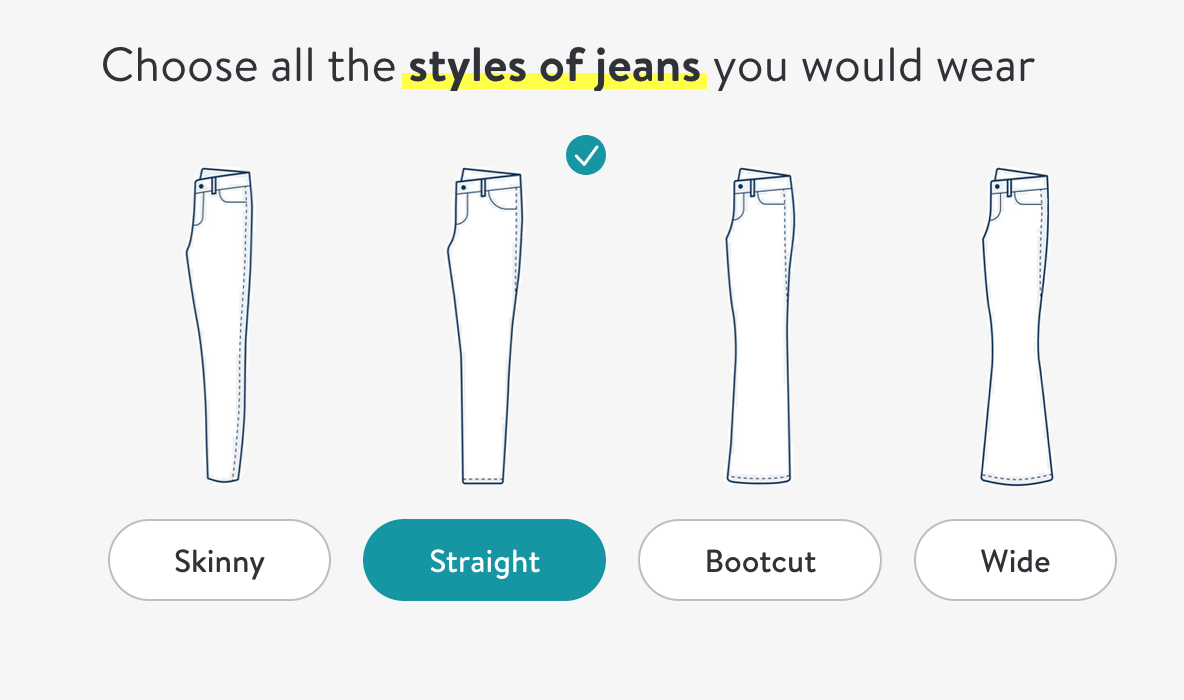} 

}

\caption{Example difference between men's (left) and women's (right) StitchFix Style Quizzes: Question on jean preferences. Contrast the abstract presentation of jeans shown to women with a picture of someone actually wearing jeans shown to men (current as of 2019-12-16).}\label{fig:stitchfix-abstraction}
\end{figure}

\begin{figure}

{\centering \includegraphics[width=0.33\linewidth]{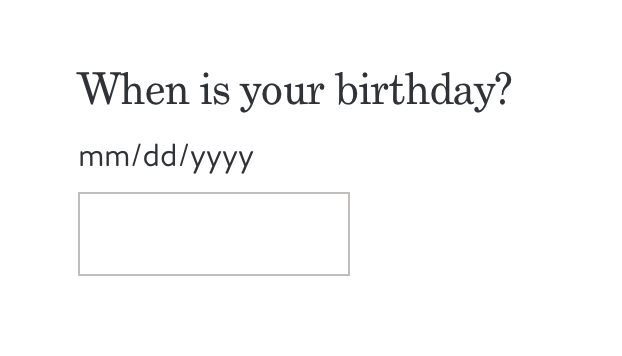} \includegraphics[width=0.33\linewidth]{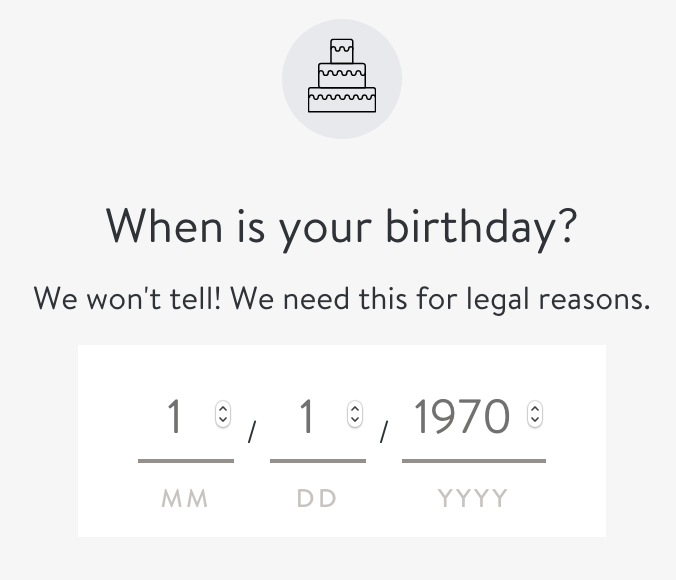} 

}

\caption{Example difference between men's (left) and women's (right) StitchFix Style Quizzes: Question about age. We note the disclaimer present for women is omitted for men (current as of 2019-12-16).}\label{fig:demographic-questions}
\end{figure}

Figure \ref{fig:stitchfix-abstraction} (current as of December 16, 2019)
presents one example relating to clothing style preferences,
specifically jean cut. The prompt in the men's quiz shows photographs of
an individual actually wearing jeans, whereas the women's quiz presents
the options in a much more abstract fashion. On top of differences
relating to clothing style and fit, many differences exist in how
demographic information is collected. Figure
\ref{fig:demographic-questions} presents an example of a question
pertaining to age. While both groups are asked the same question of
``When is your birthday?'' individuals completing the women's quiz are
primed with a ``We won't tell! We need this for legal reasons!''
statement, whereas those completing the men's are not. One has to
suspect such a difference was not coincidental, but rather reflects a
prior belief of the quiz designers as to the manner in which one should
ask about age. Other differences include questions pertaining to
parenting and occupation.

Many of these differences can be attributed to prevailing biases and
beliefs on the nature of gender and thus can serve as fertile ground for
student discussions on what algorithmic bias is. While the stakes in
purchasing clothing are lower than for criminal recidivism or
residential lending, it does present students with a clear example in
which human biases in clothing preferences influence the outcome of a
(purportedly objective) machine learning algorithm \citep{o2017weapons}.
Thus, this algorithm has the potential to reinforce these biases.

Specifically, this module can satisfy three goals. First, it provides
students with an example of algorithmic bias to which they can directly
relate. This example stands in contrast to more abstract and less
accessible examples discussed in academic readings and news media, such
as facial recognition software \citep{codedbias}. Second, it asks
students to view the statistical, mathematical, and machine learning
topics covered in class through a sociological lens, in particular
relating to the nature of gender (\citet{gebru2020race} develops this
further). Third, it gives students the opportunity to think about
statistical models in a rich, real, and realistic setting, in particular
what predictor variables are being collected and what modeling
method/technique is being used.

These three goals tie into the greater goal of imbuing students with
ethical thinking by encouraging them to think about the implications of
their model and algorithm design choices beyond the strict goal of
maximizing prediction accuracy.

\hypertarget{greys}{%
\subsection{Social networks}\label{greys}}

Perhaps in part thanks to the aptly-named Facebook movie (\emph{The
Social Network}), social networks are intuitive to students. The
relatively simple mathematical formulation of networks (i.e., graphs)
makes them easy to understand, but the complex relationships and
behaviors in such networks lead to profound research problems. Moreover,
analyzing social network data leads to thorny ethical questions.

A 300-level course on statistical analysis of social network data has as
its primary objective for students to ``learn how to answer questions by
manipulating, summarizing, visualizing, and modeling network data while
being vigilant to protect the people who are represented in that data.''
Thus, ethical concerns surrounding privacy and confidentiality are woven
directly into the main course objective.

The primary textbook is \citet{kolaczyk2014}, which provides a thorough
treatment of both the theoretical and applied aspects of social network
analysis. However, supplementary readings are especially important,
since \citet{kolaczyk2014} fails to address the many complex ethical
issues that arise for these data. We employ supplemental readings to
address data ethics on topics including:

\begin{itemize}
\tightlist
\item
  collecting social network data
\item
  informed consent for social network surveys
\item
  data identifiability and privacy in social networks
\item
  link prediction
\item
  data ethics specific to social networks
\end{itemize}

In our supplementary materials we present a module applied during the
first week of class in which we use an example from popular culture (the
television show \emph{Grey's Anatomy}) to motivate ethical issues in
social network analysis (recall \citet{burton2018teach}). It has several
goals:

\begin{itemize}
\tightlist
\item
  Prime students to always think about how the data were collected
\item
  Prime students to think about the benefits of and risks of each data
  collection / analysis / visualization, etc.
\item
  Encourage students to create their own understanding of how data
  ethics pertain to social network data as opposed to being provided
  with data ethics rules. This encourages critical thinking which can
  then be transferred to other topics and types of data.
\end{itemize}

It is especially important to introduce ethical considerations on the
first day of the course to set the tone and give students the message
that data ethics is inextricable from the rest of the content of the
course.

\hypertarget{music}{%
\subsection{Copywrited music and academic research}\label{music}}

Ethical usage of data can come into conflict with copyright law. Music
usage, for example, is heavily protected by copyright laws. The field of
Music Information Retrieval (MIR) seeks to address questions about
music, such as finding all covers of a particular song or detecting the
genre of a song. In MIR, access to music is critical to conducting
research, and that access is governed by copyright laws (which are
themselves a frequent topic when teaching tech ethics
\citep{fiesler2020we}).

Music is also a medium that has a fraught history navigating the line
between sharing and violating copyright. This history is complicated by
the power dynamics at play between recording companies and artists, and
recording companies and listeners. Today, music is often consumed
through streaming services, distorting our understanding of music
ownership. Since music is heavily protected by copyright but remains
omnipresent in our lives, conversations about data access require nuance
about ownership, sharing, and the subtleties of ethical vs.~legal
considerations.

To explore data access and copyright, we provide a module in which
students have a debate about whether the music copyright laws should be
softened for those conducting MIR research. This debate is not as simple
as whether to relax these laws, instead one side is defending the role
and purpose of copyright laws for music while the other side not only
advocates for relaxing these laws but also for \emph{how} to accomplish
this. This requirement of proposing a solution required students to hold
the responsibilities of a researcher who has broad access to data in
contrast with the ease at which we can share music (and data).

Understanding that the goal of copyright is to protect artists, and then
contrasting students' experiences of accessing and digesting music, this
debate's overarching goal is to have students navigate legal
considerations (i.e., copyright) and ethical considerations (i.e., when
to share or not share data) in the contexts of pushing research forward
and of capitalist motivations of the music industry. The legal
restrictions of copyright and the ethical responsibilities of a
researcher to protect and appropriately use (and share) their data
provide a fascinating grey area in which to have this debate. The
generational experience of our current students informs their notions of
morality and access, which in turn leads them to confront legal
restrictions in an interesting way that differs from previous
generations.

This debate activity was originally part of a senior seminar introducing
students to the field of MIR, but this activity could be done in any
course where data provenance, data usage, or data access is discussed.
Students were randomly assigned to one side of the debate. In
preparation for the debate, students were required to submit a position
paper (due just before the debate) that presented a coherent argument
that is well supported by the literature. Students were also barred from
sharing arguments with each other (even if assigned to the same side of
the debate). However, they could share resources with each other (just
not their opinion of these resources).

This structure of a preparatory paper followed by a debate required
students to engage with the research process at a deep level. For the
actual debate, each side was given opportunities to present their ideas
and offer rebuttals to the other side. This meant that not only did they
have to find resources and digest them, they had to discuss the ideas
both in written text and orally in a debate setting. Requiring students
to engage in this kind of ``perspective-taking'' may be valuable in its
own right \citep{muradova2021seeing, giroux2016perspective}.

\hypertarget{race}{%
\subsection{Teaching about race and ethnicity data}\label{race}}

In an upper-level research seminar on intergroup relationships
cross-listed in the psychology department, students learn the psychology
of close relationships between people who have differing social group
identities (e.g., racial/ethic and gender group identities). In
addition, students learn to analyze dyadic data through multilevel
modeling (i.e., mixed linear modeling), and write reproducible research
reports in APA format with the R package \texttt{papaja}
\citep{R-papaja}. This course attracts a diverse group of students in
terms of majors, professional goals, interests, statistical preparation,
and personal identities. In this ethics module, we describe a discussion
and data cleaning activity used to get students thinking in a more
careful and nuanced way about the use of race and ethnicity data in
their analyses.

The instructor provides psychological data from her own research
program, and the overarching focus of the course is to form research
questions answerable through the analysis of data that has already been
collected. Since the focus is on analyzing existing data (in addition to
talking about race), we also discuss:

\begin{itemize}
\tightlist
\item
  how to transparently communicate one's use of confirmatory versus
  exploratory analyses
\item
  the philosophical differences between inductive and deductive
  reasoning
\item
  the prevention of p-hacking
  \citep[\citet{wasserstein2019moving}]{wasserstein2016asa} and HARKing
  {[}Hypothesizing After the Results are Known;
  \citet{kerr1998harking}{]}
\end{itemize}

On the first day of this course, we have a class discussion about how we
will try to create a climate of psychological safety
\citep{edmondson1999psychological} together. This initial discussion
helps to set the tone of respect and generosity that we will need in
order to have fruitful discussions about race and ethnicity data. In the
first half of the course, class sessions alternate between discussions
about assigned readings (from psychology) and the statistical and data
science instruction they need to complete their projects. In the second
half of the course, class sessions are mainly used for actively working
on their projects. The two parts of this ethics module (discussion and
data cleaning) might be split across two class sessions.

The activity described in this module consists of a class discussion
about race and a race/ethnicity data cleaning activity in the context of
a psychology article about interracial roommate contact
\citep{shook2008interracial}. The structure of this activity invites
students to discuss the article first in small groups, and then as a
class. The larger class discussion portion of this activity is designed
to evolve into a broader discussion about the coding and use of race and
ethnicity data in quantitative research. Some important revelations that
might be pulled from the discussion include:

\begin{itemize}
\tightlist
\item
  Researchers studying interracial interactions make choices about who
  to focus on, and, in the past, this choice has often been to focus on
  white participants only. An acknowledgement of white privilege and
  who, historically, has been asking the research questions might come
  out as well.
\item
  A person's personal racial/ethnic identity may be different from how
  they are perceived by another person (roommate).
\item
  The choice to use a person's own racial/ethnic identity data or
  someone's perception of their race depends, in part, on the research
  question. When is identity or perception more important for the
  specific research context?
\item
  Race is not as clear of a categorical variable as we think it is. Can
  we think of other instances of this, for example, with gender
  categorization?
\item
  Are there times when it could serve a social good to use race in our
  analyses and, in contrast, are there ways in which using race and
  ethnicity data in analyses might reify socially constructed racial
  categories?
\item
  If you decide to use race in your analyses, what might you do in
  smaller samples if there are very small numbers of ethnic minority
  groups relative to White/European-Americans? Is it ever OK to collapse
  racial/ethnic categories? What immediate consequences do these choices
  have for the interpretation of your analysis and what broader
  consequences might these choices have when your results are consumed
  by your intended audience?
\end{itemize}

The second part of this activity asks students to code raw
race/ethnicity data into a new categorical variable called
\texttt{race\_clean}. They do this part in pairs. Then, in small groups,
they discuss the decisions they made when completing this task and also
any feelings they had during the task, as those feelings reflect the
hard realities that researchers must confront in their work. The raw
data comes in check-all-that-apply and free response formats. Students
will find this task quite difficult, and perhaps uncomfortable. The goal
is not to have them finish, but to get them to recognize the ambiguity
inherent the construction of categorical race/ethnicity variables. They
may have used the clean version of race/ethnicity variables in the past
without thinking much of it.

Lastly, the module also contains notes on closing thoughts the
instructor might offer their students after this activity. It is very
important not to skip the wrap-up for this activity. Let students know
that this is not the end of the discussion. As future data scientists,
they can play an active role in creating ethical guidelines for moving
towards more appropriate use of race and ethnicity data---a critical
area of need \citep{gebru2020race, benjamin2019race, d4bl, codedbias}.

\hypertarget{capstone}{%
\subsection{\texorpdfstring{\emph{Weapons of Math Destruction} in the
senior
capstone}{Weapons of Math Destruction in the senior capstone}}\label{capstone}}

In the senior capstone course, roughly 25\% of the course is devoted to
learning about data science ethics. During the first half of the
semester, we spend \emph{every other class period} discussing ethical
considerations that arise from weekly readings of \citet{o2017weapons}.
These readings introduce students to episodes in which often
well-intentioned data science products have had harmful effects on
society (e.g., criminal sentencing algorithms, public school teacher
evaluations, US News and World report college rankings, etc.). These
episodes are accessible to students and provide many opportunities to
engage students in thoughtful conversation.

The material in \citet{o2017weapons} also intersects with a wide variety
of statistical topics, such as modeling, validation, optimization,
Bayesian statistics, A/B testing, Type I/II errors, sensitivity and
specificity, reliability and accuracy, Simpson's paradox,
multicollinearity, confounding, and decision trees. A clever instructor
could probably build a successful course entirely around these topics.

Moreover, the ethical considerations that \citet{o2017weapons} raises
about algorithmic bias, informed consent, transparency, and privacy,
also touch on hot-button social questions surrounding structural racism,
gender equity, software licensing, cheating, income inequality,
propaganda, fake news, scams, fraud, pseudoscience, and policing bias.
Situated in the fallout from the
\href{https://en.wikipedia.org/wiki/Financial_crisis_of_2007\%E2\%80\%932008}{2008
global financial crisis}, but presaging Cambridge Analytica and fake
news, the book feels simultaneously dated and relevant. Our students
lived through the global financial crisis but most were too young to
understand it---for many of them the book allows them to grapple with
these events for the first time as adults.

The first major goal of the module is to have students interrogate the
manifold ethical considerations in data science. Reading
\citet{o2017weapons} and in-class active learning activities help
accomplish this learning goal. We employ a variety of techniques,
including think-pair-share, breakout groups, student-led discussions,
and even lecturing to keep students engaged in class. This work helps
students achieve the low-level ``identification'' thinking from
\citet{bloom1956taxonomy}.

However, the second major goal is to have students interpret the actions
of data scientists in real-world scenarios and communicate their
evaluations in writing. To this end, more structured readings are needed
(these are the ``resources, such as professional guidelines'' that are
alluded to in our learning goal). We present students with two
frameworks for thinking critically about data science ethics: Data
Values and Principles \citep{values2019} and the Hippocratic Oath for
Data Science \citep{national2018data}. \footnote{The Oxford-Munich Code
  of Conduct for Professional Data Scientists is another similar effort:
  \url{http://www.code-of-ethics.org/code-of-conduct/}} The former
defines four values (inclusion, experimentation, accountability, and
impact) and twelve principles that ``taken together, describe the most
effective, ethical, and modern approach to data teamwork.'' The latter
provides a data science analog to the oath that medical doctors have
taken for centuries. Students then complete two assignments that require
high-level ``evaluative'' thinking (again from Bloom).

First, each student writes an essay in which they analyze a data science
episode---perhaps drawn from \citet{o2017weapons}---in the context of
one of the aforementioned frameworks. This assignment forces students to
assess whether the actions of specific data scientists were ethical,
using published resources for guidance. Second, each project group
(consisting of 3--5 students) writes an ethical statement about their
semester-long project, in which they collectively describe any ethical
issues related to their work, foresee possible negative ramifications of
their work, and justify the value of their project.

Together, these assignments not only impress upon students the
importance of ethics in data science, but also give them tools and
experience to reason constructively about data science ethics in the
future. The goal is to produce students who have fully integrated ethics
into their understanding of statistics and data science and who possess
the knowledge to evaluate actors in the field.

\hypertarget{assess}{%
\section{Assessment of our ethical curriculum}\label{assess}}

Early returns suggest that our emphasis on teaching data science ethics
is having an impact and producing graduates with the ability to
translate what they have learned into action. To support this claim we
relate five concrete anecdotes, analyze results from an anonymous
student survey, and reflect on the experience of faculty in our program.

\hypertarget{sec:action}{%
\subsection{Data science ethics in action}\label{sec:action}}

There is very little research on the impact of data science ethics
instruction on undergraduate students. Thus, we present the following
five examples, which---while anecdotal--- illustrate how our (former)
students are putting their knowledge of data science ethics into
practice. In each case, we tie their actions to the taxonomy in
\citet{bloom1956taxonomy} and our learning goal.

Two students in the senior capstone course were so engaged with the
discussion of ethics surrounding the OkCupid data from
\citet{kim2015okcupid} that they independently wrote a letter to the
editor calling for a review and recommending specific modifications
\citep{xiao2021letter}. This letter formed the basis of the correction
to the original article published in May 2021
\citep{kim2015okcupidcorrection}. While their efforts had the full
support of the authors, this was \emph{not} part of the capstone course
and they received no credit for it. Their work necessitates high-level
thinking about data science ethics in context (evaluation and synthesis)
and demonstrates attainment of our learning goal.

Another student used her experience with data science ethics directly in
a summer internship with an anonymous company to help draft the
company's heretofore non-existent policies around ethical data use
\citep{rachel2019}.

\begin{quote}
``{[}She{]} was also the first data scientist to work in the
{[}company{]} space. Until her arrival, {[}company{]}'s businesses
lacked clear guidelines for collecting data and ways for using that data
to generate insights. Surprised by this, {[}she{]} first initiated
conversations with the {[}company{]} team around ethical concerns in
data collection. \textbf{Drawing on lessons from her academic work}, and
discussions with her Smith mentors, she helped to develop policies for
{[}company{]} businesses to ethically collect, manage, and act on
customer data moving forward.''
\end{quote}

We note that the connection between data science ethics in practice and
her academic coursework was made explicit \emph{by the student}. Again,
this generative work represents high-level thinking that meets our
learning goal.

In 2021, a current SDS major won a national prize in a Human Rights
Essay Contest sponsored by the American Association for the Advancement
of Science. Although her essay was about telehealth---not normally
considered a data science topic---the student connected her success
directly to the treatment of ethics in her SDS courses \citep{solow21}:

\begin{quote}
``Something else Smith does really well is include the ethical
element--understanding how you make choices based on your own biases,''
she adds. ``Every statistics class I've had here has included a
discussion of ethics.''
\end{quote}

\noindent The student's experience supports our claim that ethics is
woven into the curriculum and her essay constitutes high-level thinking
that meets our learning goal.

\begin{figure}

{\centering \includegraphics[width=0.66\linewidth]{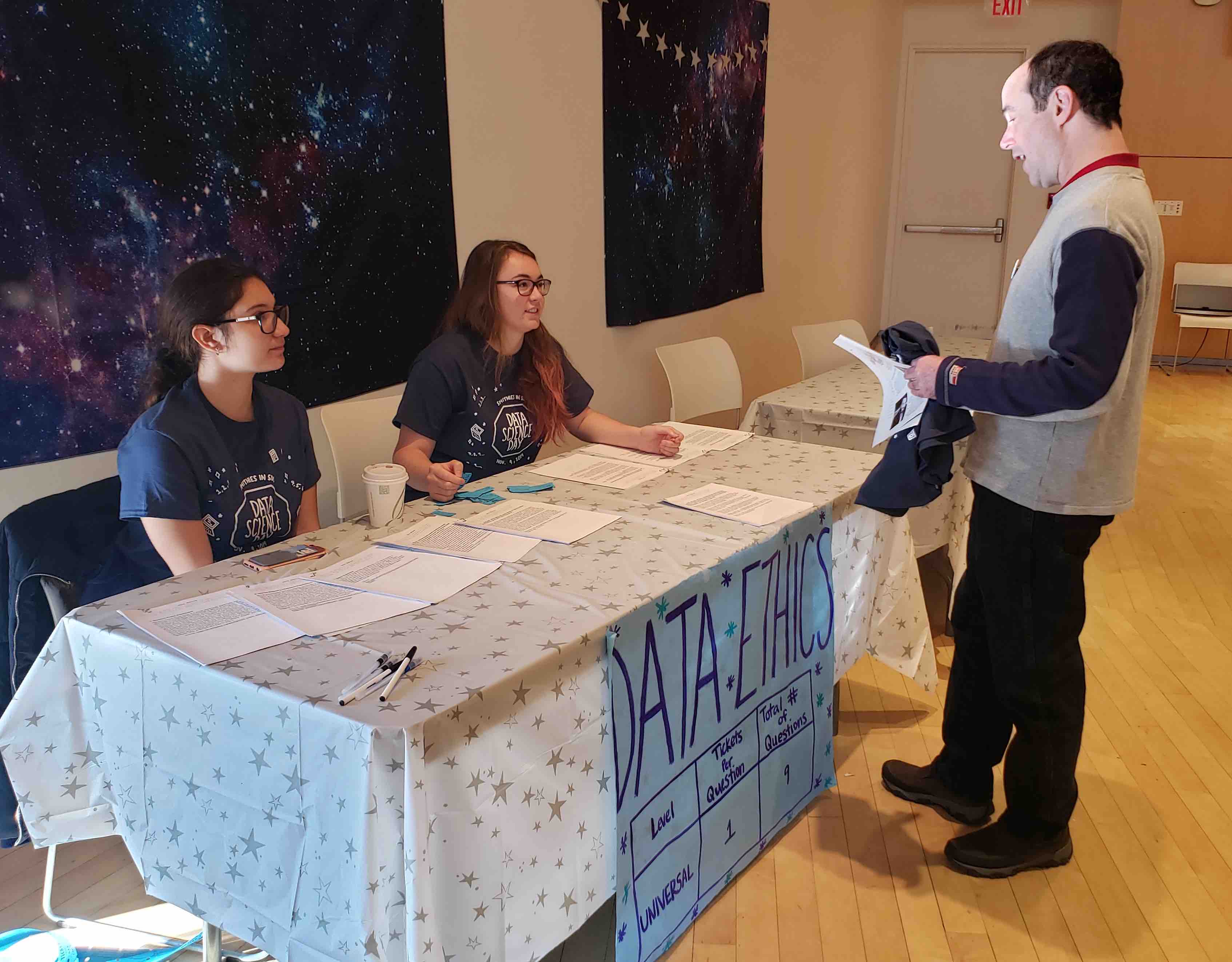} 

}

\caption{The SDS student group chose to staff a `data ethics' booth at Data Science Day 2019.}\label{fig:ssds}
\end{figure}

Another anecdote involves a student group supporting students in our
major that held their annual ``Data Science Day'' on November 9th, 2019.
At the open house portion of the event, in addition to operating booths
on data visualization and machine learning, students set up a ``data
ethics'' booth with handouts posing ethical and philosophical questions
about the use of data (see Figure \ref{fig:ssds}). While this event was
sponsored by the program, programming for the event was entirely
determined by students. The inclusion of the booth suggests that
students see ethics as an integral component of data science, on par
with data visualization and machine learning. Although these actions do
not meet our learning goal, they do reflect an ability to formulate
ethical conundrums in data science that indicates high-level thinking.

Furthermore, in the wake of discussions on racism and white supremacy
spurred by the murder of George Floyd in May 2020, two students created
a \href{https://hmsnell.github.io/DataSci-ResourcesForChange/}{Data
Science Resources for Change} website. They state: ``In order to be
thoughtful, effective, and inclusive data scientists, we believe it is
important to understand the ways in which bias can play a dangerous role
within our field, to understand the ways in which data can be used to
either reinforce/exacerbate or fight oppression, and to support the
inclusion of voices of color within the community.'' To this end this
website includes numerous resources such as reading lists, videos and
podcasts, organizations to support, and notable people to follow. These
actions reflect lower-level thinking and a powerful desire to be part of
a solution to complex problems.

We interpret all of these as early signs of our program's success at
producing more ethical data scientists.

\hypertarget{survey}{%
\subsection{Analysis of survey responses}\label{survey}}

\begin{figure}
\includegraphics[width=1\linewidth]{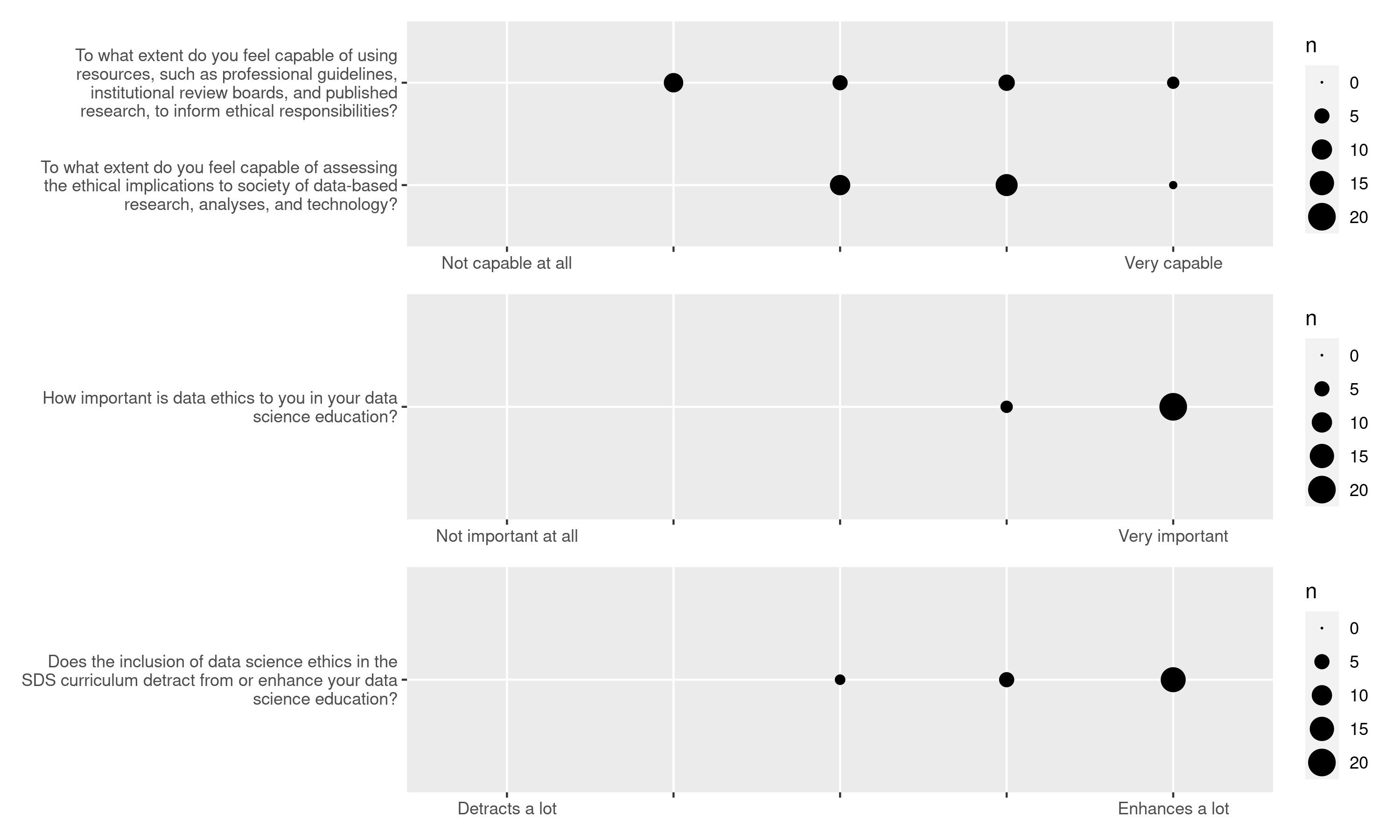} \caption{Student self-assessment of their ethical capabiliities, and the importance of data science ethics in their education, from an anonymous survey of 23 students. We note that nearly all respondents saw the inclusion of data science ethics as an important enhancement to their education, although they were less certain of their own capabilities in analyzing ethical concerns. Created using \texttt{ggplot2} \citep{R-ggplot2} for R \citep{R-base}.}\label{fig:importance}
\end{figure}

We conducted an
\href{https://docs.google.com/forms/d/1cKecLdiRA7ydAH4ccZT6m1XBNRPi_pmPUf8kcxUfqyQ/}{anonymous
online survey} during the summer of 2019, in which 23 students
participated. \footnote{This survey was approved by the Smith College
  IRB, protocol 18-111.} The results in Figure \ref{fig:importance}
reveal that students are interested in learning more about data science
ethics and feel that it is an important part of their education.
However, they are less certain that they have achieved our stated
learning goal. Unfortunately, none of the respondents had taken the
capstone course (see Section \ref{capstone}), and so these results
almost certainly undersell the effectiveness of our ethical curriculum.

The first panel in Figure \ref{fig:importance} reflects self-assessments
from students about two aspects of our major learning goal. The
questions reflect both the ability of a student to assess the ethical
implications of data science work, as well as their ability to draw on
published materials to inform their thinking. These ideas are most
explicitly and thoroughly tackled in the senior capstone, and so the
lack of respondents with that course under their belt renders this
picture incomplete.

The second panel addresses the importance of ethics to a student's data
science education. Here, students \emph{universally} believe that data
science ethics is important \emph{to them} in their education, with most
responding that it is ``very important.'' This finding supports the
recommendation of \citet{national2018data}.

Finally, the third panel in Figure \ref{fig:importance} makes plain that
no students feel that the inclusion of data science ethics detracts from
their data science education, with most students seeing the inclusion as
an enhancement. We encourage data science programs contemplating adding
ethical content to consider this point particularly. That is, the
respondents to this survey did not see the inclusion of data science
ethics as a distraction from more important, interesting, technical, or
valuable content. Rather, learning about data science ethics enhances
that curriculum.

\hypertarget{sec:reflect}{%
\subsection{Self-reflection}\label{sec:reflect}}

The faculty in SDS remains committed to our goal of developing data
scientists who can assess the ethical implications of their work. At the
same time, the greater emphasis on ethics is not without its challenges.

The emphasis on data science ethics has permeated our departmental
culture in a positive way. Because we are all attuned to the ways in
which ethics intersect with our teaching, research, and curriculum
development, discussions of ethical considerations are natural and no
one feels like they always have to be the one person to surface ethical
concerns. Raising awareness about data science ethics often means
bringing news items into the classroom, which is good practice that
helps us stay current and connect our academic work to current events.
Seeing our students put their ethical training into practice (see
Section \ref{sec:action}) is particularly rewarding. Many of our
students seem particularly drawn to ethical questions in data science,
and seeing the ways in which they are able to integrate what they learn
in our classes into their other classes, as well as their personal and
professional lives, is gratifying.

The biggest challenge for faculty is navigating our role in the field of
data science ethics. Staying current with notable ethical breaches
enables us to raise awareness among students, but doesn't make us
experts in the field of data science ethics. Most of us do not consider
data science ethics to be among our fields of research---do we possess
the knowledge to teach these topics at sufficient depth? As noted in
Section \ref{approaches}, most of us have little to no formal training
in applied ethics---are we capable of helping students reason about why
certain actions are ethical or unethical? For the most part, our
training in the humanities is at the undergraduate level---how well do
we assess the critical thinking of our students in essay form? These
questions reflect (what we hope is a healthy) academic tension between
``our careers as we imagined them in graduate school'' and ``our careers
as we think they should be now.'' In Section \ref{sec:next}, we provide
some greater institutional context for these questions.

\hypertarget{conclusion}{%
\section{Conclusion}\label{conclusion}}

The landscape of data science ethics continues to evolve. We conclude
with some next steps, institutional considerations, and final thoughts.

\hypertarget{sec:next}{%
\subsection{Next steps and institutional context}\label{sec:next}}

A central challenge noted in Sections \ref{approaches} and
\ref{sec:reflect} is the lack of formal training in ethics among the SDS
faculty. We are pursuing several avenues to improve the richness of our
ethical curriculum, taking advantage of our liberal arts setting where
possible.

First, while changing the expertise of the faculty is a long-term
process, bringing considerations of data science ethics into our hiring
practices has already been helpful (see Section \ref{program}). Our most
recent tenure-track hire has expertise in data ethnography that both
broadens the scope of the ethical questions we can present to students,
and perhaps more importantly, significantly increases the depth to which
students can pursue their interest in data science ethics. This person
is offering a new course in data ethnography and a first-year seminar
(FYS 189) on intersections between data and social justice.
(\citet{elisa2020data} discusses the relationship between social justice
and related courses on data literacy.)

Second, we are currently exploring potential points of intersection
between SDS and the philosophy department. While the minor in ethics at
Smith was recently decommissioned (due to unreplaced retirements among
faculty with scholarly expertise \emph{in ethics}), those who remain
have increasingly directed their courses towards applied ethics, often
in the context of data. A recent first-year seminar (FYS 105) is titled
``Ethics of Big Data.'' These first-year seminars---which all entering
students are required to take as part of our liberal arts curriculum and
which are often interdisciplinary---are a useful mechanism for helping
students---who might not otherwise have a fully-developed interest in
data science---connect data science to larger issues in society. A
standalone course on data science ethics, possibly co-taught by members
of both SDS and philosophy, is another possible innovation under
discussion.

Third, at the same time that the ethics program is closing, interest in
ethics among scientists is only increasing. A groups of chemists and
biologists are designing a course called ``Ethics and Scientific
Research.'' Our colleague in computer science is teaching a course
called ``Responsible Computing'' (recall the Mozilla Responsible
Computer Science Challenge). A colleague in government teaches a course
called ``The Politics of Data'' that discusses
\citet{zuboff2018surveillance} and counts towards the communication
requirement for the SDS major. The proximity to data science-adjacent
scholars, as well as philosophers trained in applied ethics, is one
advantage of reforming curriculum in a liberal arts environment such as
Smith. Leveraging these efforts to improve our own major is something we
are actively exploring.

Finally, more rigorous assessment of our efforts is necessary, as we
build towards a curricular mapping exercise (this coming year) and a
decennial review (in three years).

\hypertarget{portable}{%
\subsection{Portability}\label{portable}}

While all institutions are unique, we have designed our ethical modules
to be portable. Any instructor should be able to put our modules into
practice in their own classroom with a minimal amount of customization
and preparation. Some of the departmental initiatives described in
Section \ref{dept} require money, curricular flexibility, or exchanges
with nearby colleges that may not exist at other institutions. However,
these initiatives merely implement the recommendations of
\citet{national2018data}, so all institutions should be able to use that
call to action to push for greater institutional support, where
necessary.

Likely the biggest obstacle to replicating our work at your institution
is the potentially large discrepancy between the student culture at
Smith (a selective, liberal arts college for women) and yours. We note
that the vast majority of authors in the data ethics space happen to be
women, many of whom established research programs in foundational
machine learning and data science before examining data and algorithmic
ethics. Researchers pushing the bounds of both technical and ethical
considerations can be important role models in a field with too many
examples of technical work being used to maintain or exacerbate
inequities. We encourage readers to consider the possible benefits of
including ethics in your data science curriculum, particularly in terms
of retaining student interest from introductory courses through senior
capstone courses.

\hypertarget{final-thoughts}{%
\subsection{Final thoughts}\label{final-thoughts}}

The long-term health of data science as a discipline relies on public
trust. Ethical lapses, or gross indifference to ethics, has resulted in
the deployment of data science products that are harmful to society, due
to biases that we now recognize. Our students are part of the generation
of data scientists who will address these issues and restore faith in
data-driven applications. In order to do this, they need to not only
recognize ethical considerations as integral to data science, but also
have the ability to assess the ethical behavior of data scientists. We
present our approach to achieving this in the hopes that others will
emulate and refine what we have started.

\bibliographystyle{agsm}
\bibliography{bibliography.bib,packages.bib}

\end{document}